\documentclass[aps,prx,floatfix,twocolumn,nofootinbib]{revtex4-1}
\pdfoutput=1
\usepackage[utf8]{inputenc}
\usepackage[T1]{fontenc}
\usepackage{physics}
\usepackage{graphicx}
\usepackage{amsmath, amsfonts, amssymb}
\usepackage{bbm, bm}
\usepackage{color}
\usepackage{hyperref}
\usepackage[ruled,vlined]{algorithm2e}
\usepackage{upgreek}
\usepackage{nicefrac}
\usepackage{fancyvrb}

\usepackage{tikz}
\usetikzlibrary{quantikz}

\DeclareMathOperator{\Arccosh}{Arccosh}
\newtheorem{theorem}{Theorem}

\begin{document}

\title{Classical variational simulation of the Quantum Approximate Optimization Algorithm}

\author{Matija Medvidović}
\affiliation{Center for Computational Quantum Physics, Flatiron Institute, 162 5th Avenue, New York, NY 10010, USA}
\affiliation{Department of Physics, Columbia University, New York 10027, USA}

\author{Giuseppe Carleo}
\affiliation{Institute of Physics, École Polytechnique Fédérale de Lausanne (EPFL), CH-1015 Lausanne, Switzerland}

\begin{abstract}
    A key open question in quantum computing is whether quantum algorithms can potentially offer a significant advantage over classical algorithms for tasks of practical interest. Understanding the limits of classical computing in simulating quantum systems is an important component of addressing this question. We introduce a method to simulate layered quantum circuits consisting of parametrized gates, an architecture behind many variational quantum algorithms suitable for near-term quantum computers. A neural-network parametrization of the many-qubit wave function is used, focusing on states relevant for the Quantum Approximate Optimization Algorithm (QAOA). For the largest circuits simulated, we reach 54 qubits at 4 QAOA layers, approximately implementing 324 RZZ gates and 216 RX gates without requiring large-scale computational resources. For larger systems, our approach can be used to provide accurate QAOA simulations at previously unexplored parameter values and to benchmark the next generation of experiments in the Noisy Intermediate-Scale Quantum (NISQ) era.
\end{abstract}

\maketitle

\section{Introduction}

The past decade has seen a fast development of quantum technologies and the achievement of an unprecedented level of control in quantum hardware \cite{Arute2019}, clearing the way for demonstrations of quantum computing applications for practical uses. However, near-term applications face some of the limitations intrinsic to the current generation of quantum computers, often referred to as Noisy Intermediate-Scale Quantum (NISQ) hardware \cite{preskill_quantum_2018}. In this regime, a limited qubit count and absence of quantum error correction constrain the kind of applications that can be successfully realized. Despite these limitations, hybrid classical-quantum algorithms \cite{Peruzzo2014, Farhi2018, Farhi2014, Grant2018} have been identified as the ideal candidates to assess the first possible advantage of quantum computing in practical applications \cite{Aspuru-Guzik2005, OMalley2016, Biamonte2017, Lloyd1996}.

The Quantum Approximate Optimization Algorithm (QAOA) \cite{Farhi2014} is a notable example of variational quantum algorithm with prospects of quantum speedup on near-term devices. Devised to take advantage of quantum effects to solve combinatorial optimization problems, it has been extensively theoretically characterized \cite{Wang2018, Farhi2014_2, Lloyd2018, Jiang2017, Hadfield2019, zhou_quantum_2020}, and also experimentally realized on state-of-the-art NISQ hardware \cite{Arute2020}. While the general presence of quantum advantage in quantum optimization algorithms remains an open question \cite{santoro_theory_2002, ronnow_defining_2014, Guerreschi2019, Bravyi2020}, QAOA has gained popularity as a quantum hardware benchmark \cite{Pagano2019, Bengtsson2019, Willsch2020, Otterbach2017}. As its desired output is essentially a classical state, the question arises whether a specialized classical algorithm can efficiently simulate it \cite{Bravyi2019}, at least near the variational optimum. 
In this paper, we use a variational parametrization of the many-qubit state based on Neural Network Quantum States (NQS) \cite{Carleo2017} and extend the method of Ref.~\cite{Jonsson2018} to simulate QAOA. This approach trades the need for exact \textit{brute force} exponentially scaling classical simulation with an approximate, yet accurate, classical variational description of the quantum circuit. In turn, we obtain an heuristic classical method that can significantly expand the possibilities to simulate NISQ-era quantum optimization algorithms. We successfully simulate the Max-Cut QAOA circuit \cite{Farhi2014, Arute2020, Wang2018} for $54$ qubits at depth $p=4$ and use the method to perform a variational parameter sweep on a 1D cut of the parameter space. The method is contrasted with state-of-the-art classical simulations based on low-rank Clifford group decompositions \cite{Bravyi2019}, whose complexity is exponential in the number of non-Clifford gates as well as tensor-based approaches \cite{Villalonga2020}. Instead, limitations of the approach are discussed in terms of the QAOA parameter space and its relation to different initializations of the stochastic optimization method used in this work.

\section{Results}

\begin{figure*}
    \centering
    \begin{quantikz}
        \lstick[wires=5, label style={xshift=-12pt, anchor=center, rotate=90}]{\Large $\ket{\psi _{\theta =0}} \propto \ket{+}$} & \gate[wires=5, style={fill=blue!15}]{U_C (\gamma _1)} \gategroup[wires=5,steps=1,style={draw=none, dashed, rounded corners, inner xsep=2pt}, background,label style={label position=below, anchor=north, yshift=-0.2cm}]{Exact} \gategroup[wires=5,steps=1,style={draw=none, dashed, rounded corners, inner xsep=2pt}, background,label style={label position=above, anchor=north, yshift=0.3cm}]{$\displaystyle  N_\text{hid} \rightarrow | E | $} & \gate[style={fill=yellow!20}]{RX(\beta_1)} \gategroup[wires=5,steps=1,style={dashed, rounded corners, inner xsep=2pt}, background,label style={label position=below, anchor=north, yshift=-0.2cm}]{Approx.} & \gate[wires=5, style={fill=blue!15}]{U_C (\gamma _2)} \gategroup[wires=5,steps=1,style={draw=none, dashed, rounded corners, inner xsep=2pt}, background,label style={label position=below, anchor=north, yshift=-0.2cm}]{Exact} \gategroup[wires=5,steps=1,style={draw=none, dashed, rounded corners, inner xsep=2pt}, background,label style={label position=above, anchor=north, yshift=1.1cm}]{\Large $\displaystyle \substack{ N_\text{hid} \\ \downarrow \\ 2 N_\text{hid} }$} & \qw \slice[style=red, label style={inner sep=8pt}]{\Large $\displaystyle \substack{ N_\text{hid} \\ \downarrow \\ \nicefrac{N_\text{hid}}{2} }$} & \qw & \gate[style={fill=yellow!20}]{RX(\beta_2)} \gategroup[wires=5,steps=1,style={dashed, rounded corners, inner xsep=2pt}, background, label style={label position=below, anchor=north, yshift=-0.2cm}]{Approx.} & \ \ldots\ \qw & \gate[wires=5, style={fill=blue!15}]{U_C (\gamma _p)} \gategroup[wires=5,steps=1,style={draw=none, dashed, rounded corners, inner xsep=2pt}, background,label style={label position=below, anchor=north, yshift=-0.2cm}]{Exact} \gategroup[wires=5,steps=1,style={draw=none, dashed, rounded corners, inner xsep=2pt}, background,label style={label position=above, anchor=north, yshift=1.1cm}]{\Large $\displaystyle \substack{ N_\text{hid} \\ \downarrow \\ 2 N_\text{hid} }$} & \qw \slice[style=red, label style={inner sep=8pt}]{\Large $\displaystyle \substack{ N_\text{hid} \\ \downarrow \\ \nicefrac{N_\text{hid}}{2} }$} & \qw & \gate[style={fill=yellow!20}]{RX(\beta_p)} \gategroup[wires=5,steps=1,style={dashed, rounded corners, inner xsep=2pt}, background, label style={label position=below, anchor=north, yshift=-0.2cm}]{Approx.} & \rstick[wires=5, label style={xshift=12pt, anchor=center, rotate=90}]{\large $\ket{\psi _{\theta _\text{opt}}} \propto \ket{\bm{\gamma}, \bm{\beta}}$} \qw \\
        & & \gate[style={fill=yellow!20}]{RX(\beta_1)} & & \qw & \qw & \gate[style={fill=yellow!20}]{RX(\beta_2)} & \ \ldots\ \qw & & \qw & \qw & \gate[style={fill=yellow!20}]{RX(\beta_p)} & \qw & \\
        & & \gate[style={fill=yellow!20}]{RX(\beta_1)} & & \qw & \qw & \gate[style={fill=yellow!20}]{RX(\beta_2)} & \ \ldots\ \qw & & \qw & \qw & \gate[style={fill=yellow!20}]{RX(\beta_p)} & \qw & \\
        & & \gate[style={fill=yellow!20}]{RX(\beta_1)} & & \qw & \qw & \gate[style={fill=yellow!20}]{RX(\beta_2)} & \ \ldots\ \qw & & \qw & \qw & \gate[style={fill=yellow!20}]{RX(\beta_p)} & \qw & \\
        & & \gate[style={fill=yellow!20}]{RX(\beta_1)} & & \qw & \qw & \gate[style={fill=yellow!20}]{RX(\beta_2)} & \ \ldots\ \qw & & \qw & \qw & \gate[style={fill=yellow!20}]{RX(\beta_p)} & \qw &
    \end{quantikz}
    \caption{
        \textbf{The QAOA quantum circuit.} A schematic representation of the QAOA circuit and our approach to simulating it. The input state is trivially initialized to $\ket{+}$. Next, at each $p$, the exchange of exactly ($U_C$) and approximately ($RX (\beta ) = e^{-i \beta X} $) applicable gates is labeled (see Sec.~\ref{sec:methods}). As noted in the main text, each (exact) application of the $U_C$ gate leads to an increase in the number of hidden units by $\vert E \vert$ (the number of edges in the graph). In order to keep that number constant, we "compress" the model (see Sec.~\ref{sec:methods}), indicated by red dashed lines after each $U_C$ gate. The compression is repeated at each layer after the first, halving the number of hidden units each time, immediately after doubling it with $U_C$ gates. After the final layer, the RBM is parametrized by $\theta _{\text{opt}}$, approximating the final QAOA target state $\ket{\bm{\gamma}, \bm{\beta}}$.
    }
    \label{fig:circuit}
\end{figure*}
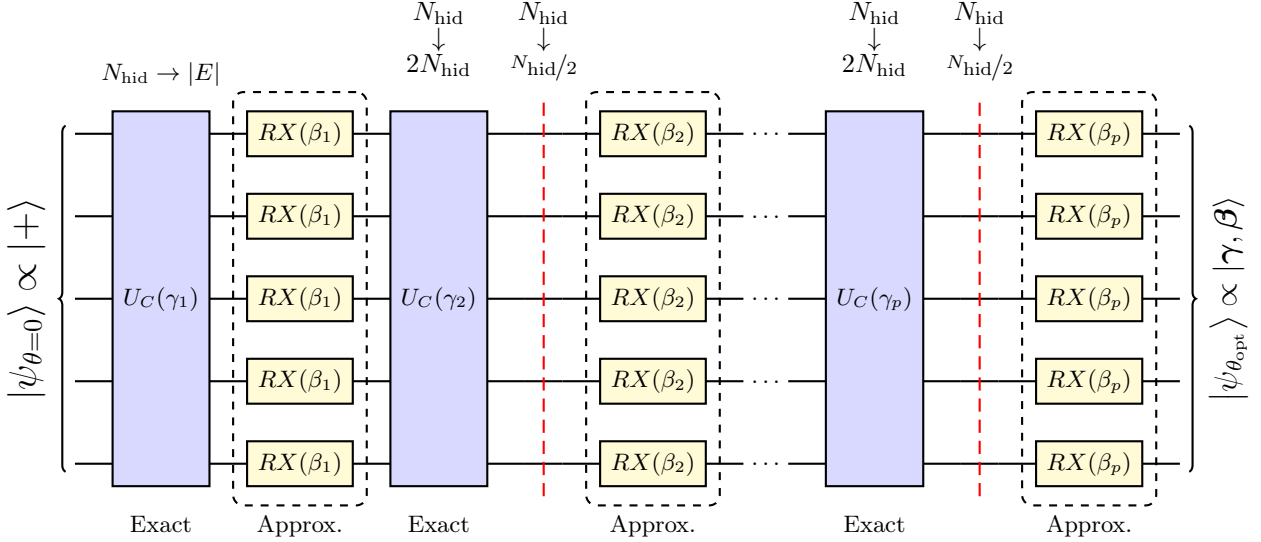

\subsection{The Quantum Approximate Optimization Algorithm}

The Quantum Approximate Optimization Algorithm (QAOA) is a variational quantum algorithm for approximately solving discrete combinatorial optimization problems. Since its inception in the seminal work of Farhi, Goldstone and Gutmann \cite{Farhi2014, Farhi2014_2}, QAOA has been applied to Maximum Cut (Max-Cut) problems. With competing classical algorithms \cite{Goemans1995} offering exact performance bounds for all graphs, an open question remains -- can QAOA perform better by increasing the number of free parameters?

In this work, we study a quadratic cost function \cite{Lucas2014, Barahona1982} associated with a Max-Cut problem. If we consider a graph $G=(V,E)$ with edges $E$ and vertices $V$, the MaxCut of the graph $G$ is defined by the following operator:
\begin{equation}
\label{eq:qaoa_cost_operator}
    \mathcal{C} = \sum _{i, j \in E} w_{i j} Z_i Z_j \; ,
\end{equation}

\noindent where $w_{ij}$ are the edge weights and $Z_i$ are Pauli operators. The classical bitstring $\mathcal{B}$ that minimizes $\bra{\mathcal{B}} \mathcal{C} \ket{\mathcal{B}}$ is the graph partition with the maximum cut. QAOA approximates such a quantum state through a quantum circuit of predefined depth $p$:
\begin{equation}
\label{eq:qaoa_output_state_def}
    \ket{\bm{\gamma}, \bm{\beta}} = U_B (\beta _p) U_C (\gamma _p) \cdots U_B (\beta _1) U_C (\gamma _1) \ket{+} \; ,
\end{equation}

\noindent where $\ket{+}$ is a symmetric superposition of all computational basis states: $\ket{+} = H^{\otimes N} \ket{0}^{\otimes N} $ for $N$ qubits. The set of $2 p$ real numbers $\gamma _i$ and $\beta _i$ for $i=1 \ldots p$ define the variational parameters to be optimized over by an external classical optimizer. The unitary gates defining the parametrized quantum circuit read $U_B (\beta) = \prod _{i \in V} e^{-i \beta X_i}$ and $U_C(\gamma) = e^{-i \gamma \mathcal{C}}$.

Optimal variational parameters $\gamma $ and $\beta$ are then found through an outer-loop classical optimizer of the following quantum expectation value:
\begin{equation}
\label{eq:qaoa_cost_general}
    C(\bm{\gamma}, \bm{\beta}) = \bra{\bm{\gamma}, \bm{\beta}} \mathcal{C} \ket{\bm{\gamma}, \bm{\beta}}
\end{equation}

It is known that, for QAOA cost operators of the general form $\mathcal{C} = \sum _k \mathcal{C} _k (Z_1, \ldots, Z_N) $, the optimal value asymptotically converges to the minimum value:
\begin{equation}
    \lim _{p \rightarrow \infty} C _p = \min _\mathcal{B} \bra{\mathcal{B}} \mathcal{C} \ket{\mathcal{B}}
\end{equation}

\noindent where $C_p$ is the optimal cost value at QAOA depth p and $\mathcal{B}$ are classical bit strings. With modern simulations and implementations still being restricted to lower $p$ values, it is unclear how large $p$ has to get in practice before QAOA becomes comparable with its classical competition.

In this work we consider 3-regular graphs with all weights $w_{i j}$ set to unity at QAOA depths of $p=1, 2, 4$.

\subsection{Classical Variational Simulation}

\begin{figure*}
    \includegraphics[width=\linewidth]{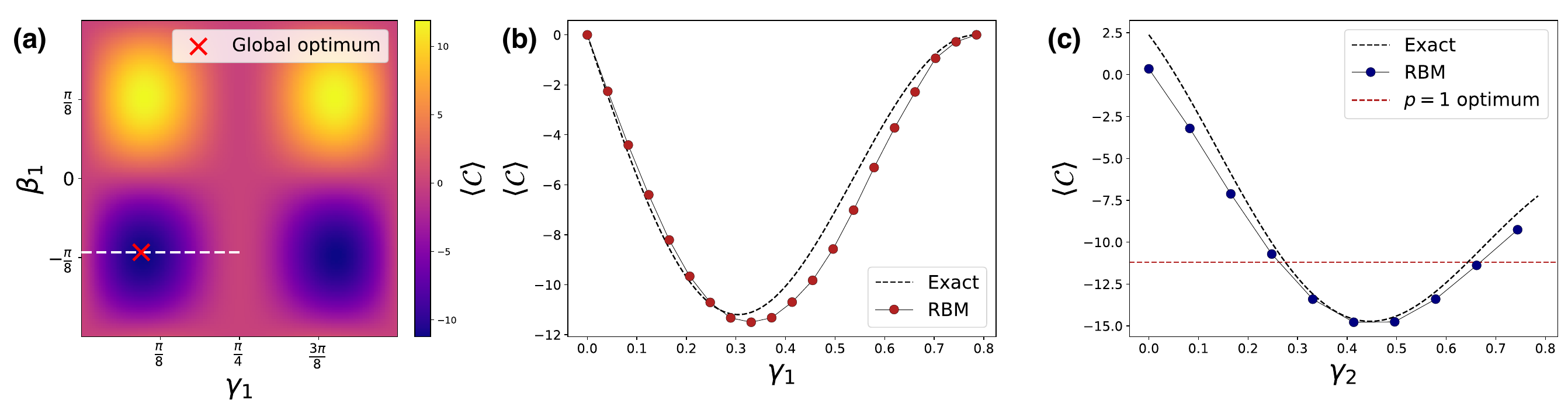}
    \caption{
        \textbf{Benchmarking the cost function for 20 qubits.}
        \textbf{a}: The exact variational QAOA landscape at $p=1$ of a random $20$-qubit instance of a 3-regular graph is presented, calculated using the analytical cost formula (see Appendix~\ref{sm:formula}). The optimum is found using a gradient-based optimizer \cite{Kingma2015} and marked. The restricted cut along the constant-$\beta$ line and at optimal $\gamma$ is more closely studied in panel b.
        \textbf{b}: RBM-based output wavefunctions are contrasted with exact results. 
        \textbf{c}: A similar variational landscape cut is presented at $p=2$. Optimal $p=2$ QAOA parameters are calculated using numerical derivatives and a gradient-based optimizer. Parameters $\gamma _1$, $\beta _1$ and $\beta _2$ are fixed at their optimal values while the cost function $\gamma _2$-dependence is investigated. We note that our approach is able to accurately reproduce the increased proximity to the combinatorial optimum associated with increasing QAOA depth $p$. (The dashed line represents the minumum from $p=1$ curve in panel b.)
    }
    \label{fig:20_qubits}
\end{figure*}

Consider a quantum system consisting of $N$ qubits. The Hilbert space is spanned by the computational basis $\{ \ket{ \mathcal{B}} : \mathcal{B} \in \{ 0,1 \} ^N \}$ of classical bit strings $\mathcal{B} = (B_1, \ldots, B_N)$. A general state can be expanded in this basis as $\ket{\psi } = \sum _{\mathcal{B}} \psi (\mathcal{B}) \ket{\mathcal{B}}$. The convention $Z_i \ket{\mathcal{B}} = (-1)^{B_i} \ket{\mathcal{B}}$ is adopted. In order to perform approximate classical simulations of the QAOA quantum circuit, we use a neural-network representation of the many-body wavefunction $\psi (\mathcal{B} )$ associated with this system, and specifically adopt a shallow network of the Restricted Boltzmann Machine (RBM) type: \cite{Hinton2002, Hinton2006, Lecun2015}
\begin{equation}
\label{eq:RBM_def}
\begin{gathered}
    \psi (\mathcal{B} ) \approx \psi _\theta (\mathcal{B} ) \equiv \exp \left( \sum _{j=1} ^{N} a_j B_j \right) \cdot \\
    \cdot \prod _{k=1} ^{N_\text{h}} \left[ 1 + \exp \left( b_k + \sum _{j=1} ^{N} W_{j k} B_j \right) \right] \; .
\end{gathered}
\end{equation}

The RBM provides a classical variational representation of the quantum state \cite{Carleo2017}. It is parametrized by a set of complex parameters $\theta  = \{ \vb{a}, \vb{b}, W \}$ -- visible biases $\vb{a} = (a_1, \ldots, a_\text{N})$, hidden biases $\vb{b} = (b_1, \ldots , b_{N_\text{h}})$ and weights $W = (W_{j, k} : j = 1 \ldots N , \; k=1 \ldots N_\text{h})$. The complex-valued ansatz given in Eq.~\ref{eq:RBM_def} is, in general, not normalized.

We note that the $N$-qubit $\ket{+}$ state required for initializing QAOA can always be exactly implemented by setting all variational parameters to $0$. That choice ensures that the wavefunction ansatz given in Eq.~\ref{eq:RBM_def} is constant across all computational basis states, as required.
The advantage of using the ansatz given in Eq.~\ref{eq:RBM_def} as an $N$-qubit state is that a subset of one- and two-qubit gates can be exactly implemented as mappings between different sets of variational parameters $\theta \mapsto \theta '$. In general, such mapping corresponding to an abstract gate $\mathcal{G}$ is found as the solution of the following nonlinear equation:
\begin{equation}
\label{param_gate_mapping}
    \braket{\mathcal{B}}{\psi _{\theta '}} = C \bra{\mathcal{B}} \mathcal{G} \ket{\psi _\theta} \; ,
\end{equation}

\noindent for all bitstrings $\mathcal{B}$ and any constant $C$, if a solution exists. For example, consider the Pauli $Z$ gate acting on qubit $i$. In that case, Eq.~\ref{param_gate_mapping} reads $e^{a' _i B_i} = C (-1)^{B_i} e^{a _i B_i}$ after trivial simplification. The solution is $a' _i = a_i + i \pi$ for $C=1$, with all other parameters remaining unchanged. 
In addition, one can exactly implement a subset of two-qubit gates by introducing an additional hidden unit coupled only to the two qubits in question. Labeling the new unit by $c$, we can implement the $RZZ$ gate relevant for QAOA. The gate is given as $RZZ(\phi) = e^{-i \phi Z_i Z_j} \propto \text{diag} (1, e^{i\phi}, e^{i\phi}, 1)$ up to a global phase. The replacement rules read:
\begin{equation}
\begin{gathered}
    W_{i c} = -2 \mathcal{A} (\phi) \; , \quad W_{j c} = 2 \mathcal{A} (\phi) \\
    a_i \rightarrow a_i + \mathcal{A} (\phi) \; , \quad a_j \rightarrow a_j - \mathcal{A} (\phi) \; ,
\end{gathered}
\end{equation}

\noindent where $\mathcal{A} (\phi ) = \Arccosh \left( e^{i \phi} \right) $ and $C=2$. Derivations of replacement rules for these and other common one and two-qubit gates can be found in Sec.~\ref{sec:methods}.

Not all gates can be applied through solving Eq.~\ref{param_gate_mapping}. Most notably, gates that form superpositions belong in this category, including $U_B (\beta) = \prod _i e^{-i\beta X_i}$ required for running QAOA. This happens simply because a linear combination of two or more RBMs cannot be exactly represented by a single new RBM through a simple variational parameter change. To simulate those gates, we employ a variational stochastic optimization scheme.

We take $\mathcal{D} (\phi, \psi) = 1 - F(\phi, \psi)$ as a measure of distance between two arbitrary quantum states $\ket{\phi}$ and $\ket{\psi}$, where $F(\phi, \psi)$ is the usual quantum fidelity:
\begin{equation}
\label{eq:fidelity}
    F(\phi, \psi) = \frac{\vert \braket{\phi}{\psi} \vert ^2}{\braket{\phi}{\phi} \braket{\psi}{\psi}} \; .
\end{equation}

In order to find variational parameters $\theta$ which approximate a target state $\ket{\phi}$ well ($\ket{\psi _\theta} \approx \ket{\phi}$, up to a normalization constant), we minimize $\mathcal{D} (\psi _\theta, \phi)$ using a gradient-based optimizer. In this work we use the Stochastic Reconfiguration (SR) \cite{Sorella1998, Metropolis1953, Hastings1970} algorithm to achieve that goal.

For larger $p$, extra hidden units introduced when applying $U_C(\gamma )$ at each layer can result in a large number of associated parameters to optimize over that are not strictly required for accurate output state approximations. So to keep the parameter count in check, we insert a model \textit{compression} step which halves the number of hidden units immediately after applying $U_C$ doubles it. Specifically we create an RBM with fewer hidden units and fit it to the output distribution of the larger RBM (output of $U_C$). Exact circuit placement of compression steps are shown on Fig.~\ref{fig:circuit} and details are provided in Sec.~\ref{sec:methods}. As a result of the compression step, we are able to keep the number of hidden units in our RBM ansatz constant, explicitly controlling the variational parameter count.

\subsection{Simulation results for 20 qubits}

In this section we present our simulation results for Max-Cut QAOA on random regular graphs of order $N$ \cite{Steger1999, Kim2003, Hagberg2008}. In addition, we discuss model limitations and its relation to current state-of-the-art simulations.

QAOA angles $\bm{\gamma}$, $\bm{\beta}$ are required as an input of our RBM-based simulator. At $p=1$, we base our parameter choices on the position of global optimum that can be computed exactly (see Appendix~\ref{sm:formula}). For $p>1$, we resort to direct numerical evaluation of the cost function as given in Eq.~\ref{eq:qaoa_cost_operator} from either the complete state vector of the system (number of qubits permitting) or from importance-sampling the output state as represented by a RBM. For all $p$, we find the optimal angles using Adam \cite{Kingma2015} with either exact gradients or their finite-difference approximations.

We begin by studying the performance of our approach on a 20-qubit system corresponding to the Max-Cut problem on a 3-regular graph of order $N=20$. In that case, access to exact numerical wavefunctions is not yet severely restricted by the number of qubits. That makes it a suitable test-case. The results can be found in Fig.~\ref{fig:20_qubits}.

In Fig.~\ref{fig:20_qubits}, we present the cost function for several values of QAOA angles, as computed by the RBM-based simulator. Each panel shows cost functions from one typical random 3-regular graph instance. We observe that cost landscapes, optimal angles and algorithm performance do not change appreciably between different random graph instances. We can see that our approach reproduces variations in the cost landscape associated with different choices of QAOA angles at both $p=1$ and $p=2$. At $p=1$, an exact formula (see Appendix~\ref{sm:formula}) is available for comparison of cost function values. We report that, at optimal angles, the overall final fidelity (overlap squared) is consistently above 94\% for all random graph instances we simulate.

\begin{figure}[!h]
    \includegraphics[width=0.98\linewidth]{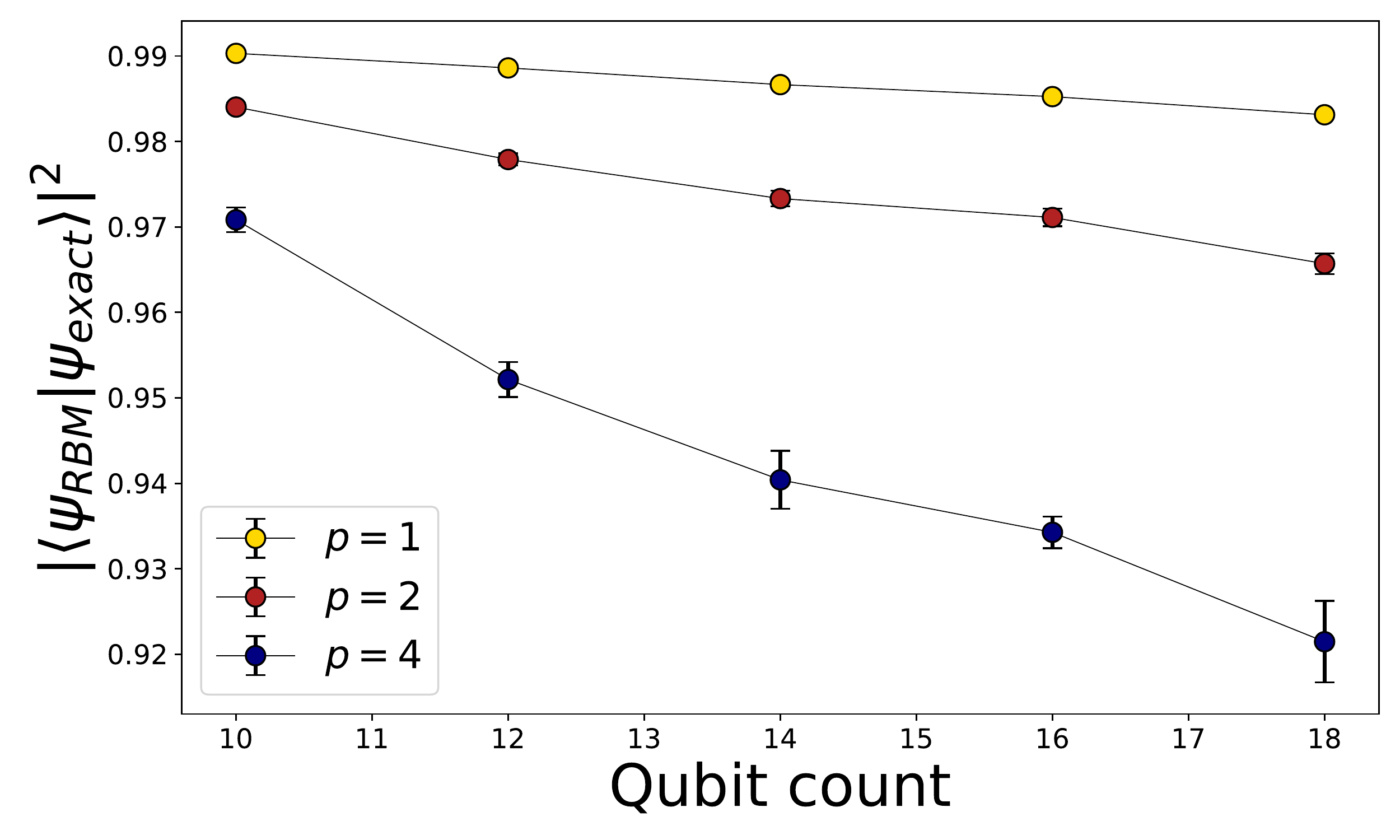}
    \caption{
        \textbf{Benchmarking with exact fidelities}. Fidelities between approximate RBM variational states and quantum states obtained with exact simulation, for different values of $p=1,2,4$ and number of qubits. QAOA angles $\bm{\gamma}, \bm{\beta}$ we use are set to optimal values at $p=4$ for each random graph instance. Ten randomly generated graphs are used for each system size. Error bars represent the standard deviation.
    }
    \label{fig:small_graph_fidelity}
\end{figure}

In addition to cost function values, we also benchmark our RBM-based approach by computing fidelities between our variational states and exact simulations. In Fig.~\ref{fig:small_graph_fidelity} we show the dependence of fidelity on the number of qubits and circuit depth $p$. While, in general, it is hard to analytically predict the behavior of these fidelities, we nonetheless remark that with relatively small NQS we can already achieve fidelities in excess of 92\% for all system sizes considered for exact benchmarks.

\subsection{Simulation results for 54 qubits}

\begin{figure*}
    \includegraphics[width=0.98\linewidth]{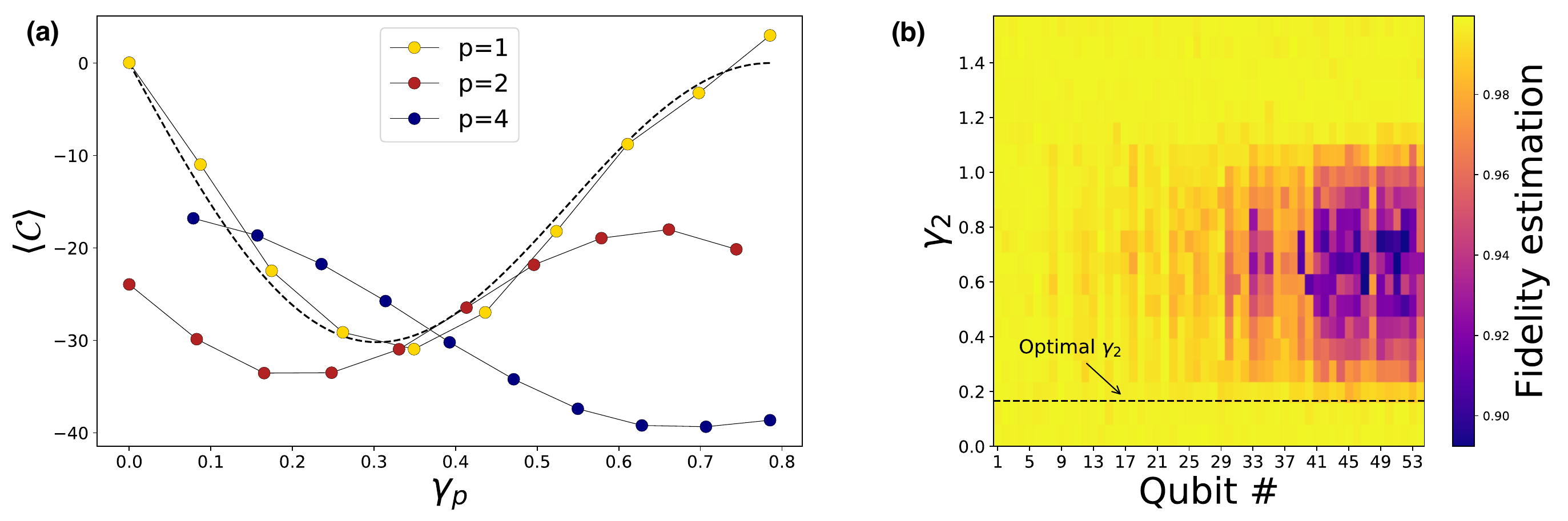}
    \caption{
        \textbf{Simulating 54 qubits.}
        \textbf{a}: Randomly generated 3-regular graphs with 54 nodes are considered at $p=1,2,4$. At each $p$, all angles were set to optimal values (except for the final $\gamma _p$) for a smaller graph of 20 nodes for which optimal angles can be found at a much smaller computational cost. Cost dependence along this 1D slice of the variational landscape (a higher-dimensional analogue of panel a of Fig.~\ref{fig:20_qubits}) is investigated. The dashed line represents the exact cost at $p=1$. Error bars were too small to be visible on the plot. At $p=2$, this 54-qubit simulation approximately implements 162 $RZZ$ gates and 108 $RX$ gates while at $p=4$ there are 324 $RZZ$s and 216 $RX$s.
        \textbf{b}: An array of final stochastic estimations of single-qubit fidelities (see Sec.~\ref{sm:approx_gate} for formula) in the course of optimizer progress. The system presented consists of 54 qubits at $p=2$ where exact state vectors are intractable for direct comparison. In these simulations, $\beta _1$, $\beta _2$ and $\gamma_1$ are kept at their optimal values. Optimal $\gamma _2$ value (for given $\beta _1$, $\beta _2$, $\gamma_1$) is shown with a dashed line. We note that the fidelity estimates begin to drop approximately as $\gamma _2$ increases beyond the optimal value. A similar qubit-by-qubit trend can be noticed across all system sizes and depths $p$ we studied.
    }
\label{fig:54qubits}
\end{figure*}

Our approach can be readily extended to system sizes that are not easily amenable to exact classical simulation. To show this, in Fig.~\ref{fig:54qubits} we show the case of $N=54$ qubits. This number of qubits corresponds, for example, to what implemented by Google's Sycamore processor, while our approach shares no other implementation details with that specific platform. For the system of $N=54$ qubits, we closely reproduce the exact error curve (see Appendix~\ref{sm:formula}) at $p=1$, implementing 81 RZZ ($e^{-i \gamma Z \otimes Z}$) gates exactly and 54 $RX$ ($e^{-i \beta X}$) gates approximately, using the described optimization method. We also perform simulations at $p=2$ and $p=4$ and obtain corresponding approximate QAOA cost function values.

At $p=4$, we exactly implement 324 $RZZ$ gates and approximately implement 216 $RX$ gates. This circuit size and depth is such that there is no available experimental or numerically exact result to compare against. The accuracy of our approach can nonetheless be quantified using intermediate variational fidelity estimates. These fidelities are exactly the cost functions (see Sec.~\ref{sec:methods}) we optimize, separately for each qubit. In Fig.~\ref{fig:54qubits} (panel b) we show the optimal variational fidelities (see Eq.~\ref{eq:fidelity}) found when approximating the action of $RX$ gates with the RBM wave function. At optimal $\gamma _4$ (minimum of $p=4$ curve at Fig.~\ref{fig:54qubits}, panel a), the lowest variational fidelity reached was above 98\%, for a typical random graph instance shown at Fig.~\ref{fig:54qubits}. As noted earlier, exact final states of 54-qubit systems are intractable so we are unable to report or estimate the full many-qubit fidelity benchmark results.

We remark that the stochastic optimization performance is sensitive to choices of QAOA angles away from optimum (see Fig.~\ref{fig:54qubits} right). In general, we report that the fidelity between the RBM state (Eq.~\ref{eq:RBM_def}) and the exact $N$-qubit state (Eq.~\ref{eq:qaoa_output_state_def}) decreases as one departs from optimal by changing $\bm{\gamma}$ and $\bm{\beta}$.

For larger values of QAOA angles, the associated optimization procedure is more difficult to perform, resulting in a lower fidelity (see the dark patch in Fig.~\ref{fig:54qubits}, panel b). We find that optimal angles were always small enough not to be in the low-performance region. Therefore, this model is less accurate when studying QAOA states away from the variational optimum. However, even in regions with lowest fidelities, RBM-based QAOA states are able to approximate cost well, as can be seen in Fig.~\ref{fig:20_qubits} and Fig.~\ref{fig:54qubits}.

As an additional hint to the high quality of the variational approximation, we capture the QAOA approximation of the actual combinatorial optimum. A tight upper bound on that optimum was calculated to be $C_\text{opt} = -69 $ for 54 qubits by directly optimizing an RBM to represent the ground state of the cost operator defined in Eq.~\ref{eq:qaoa_cost_general}.

\subsection{Comparison with other methods}

In modern sum-over-Cliffords/Metropolis simulators, computational complexity grows exponentially with the number of non-Clifford gates. With the $RZZ$ gate being a non-Clifford operation, even our 20-qubit toy example, exactly implementing 60 $RZZ$ gates at $p=2$, is approaching the limit of what those simulators can do \cite{Bravyi2019}. In addition, that limit is greatly exceeded by the larger, 54-qubit system we study next, implementing 162 $RZZ$ gates. State-of-the-art tensor-based approaches \cite{Villalonga2020} have been used to simulate larger circuits but are ineffective in the case of non-planar graphs.

\begin{figure}[h]
    \includegraphics[width=0.98\linewidth]{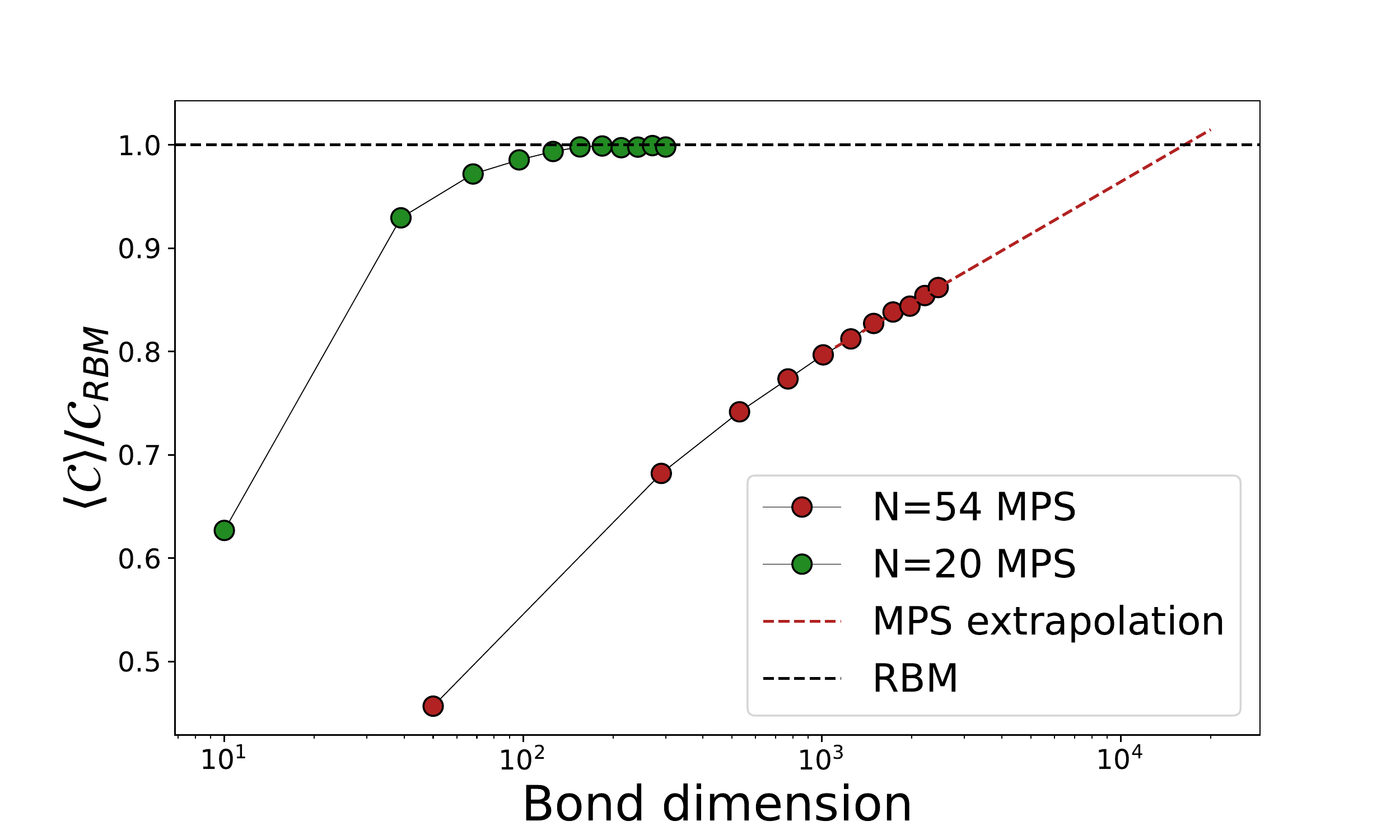}
    \caption{
        \textbf{Comparison with Matrix Product States.}
        A range of MPS-based QAOA simulations are compared to our RBM ansatz performance on both 20-qubit and 54-qubit graphs at $p=2$. In the 20-qubit case, we see quick convergence to the QAOA cost optimum with increasing bond dimension. Approximation ratio with of the RBM output is shown on the y-axis. However, on a 54-qubit graph, MPS accuracy increases approximately logarithmically with bond dimension. An approximation of the MPS bond dimension required for reaching RBM performance is extrapolated to be $\approx 1.5 \times 10^{4}$ which amounts to $\sim 10^{10}$ free parameters.
    }
    \label{fig:mps}
\end{figure}

Another very important tensor-based method is the Matrix Product State (MPS) variational representation of the many-qubit state. This is is a low-entanglement representation of quantum states, whose accuracy is controlled by the so-called bond-dimension. Routinely adopted to simulate ground states of one-dimensional systems with high accuracy \cite{White1992, Vidal2003, Vidal2004}, extensions of this approach to simulate challenging circuits have also been recently put forward \cite{zhou_what_2020}. In Fig.~\ref{fig:mps}, our approach is compared with an MPS ansatz. We establish that for small systems, MPS provides reliable results with relatively small bond dimensions. For larger systems, however, our approach significantly outperforms MPS-based circuit simulation methods both in terms of memory requirements (fewer parameters) and overall runtime.  This is to be expected in terms of entanglement capacity of MPS wave functions, that are not specifically optimized to handle non-one dimensional interaction graphs, as in this specific case at hand.

For a more direct comparison, we estimate the MPS bond dimension required for reaching RBM performance at $p=2$ and 54 qubits to be $\sim 10^4$ (see Fig.~\ref{fig:mps}), amounting to $\sim 10^{10}$ complex parameters ($\approx 160$ GB of storage) while our RBM approach uses $\approx 4500$ parameters ($\approx 70$ kB of storage). In addition, we expect the MPS number of parameters to grow with depth $p$ because of additional entanglement, while RBM sizes heuristically scale weakly (constant in our simulations) with $p$ and can be controlled mid-simulation using our compression step. It should be noted that the output MPS bond dimension depends on the specific implementation of the MPS simulator, namely, qubit ordering and the number of "swap" gates applied to correct for the non-planar nature of the underlying graph, and that a more efficient implementations might be found.  However, determining the optimal implementation is itself a difficult problem and, given the entanglement of a generic circuit we simulate, it would likely produce a model with orders of magnitude more parameters than a RBM-based approach.

\section{Discussion}

In this work, we introduce a classical variational method for simulating QAOA, a hybrid quantum-classical approach for solving combinatorial optimizations with prospects of quantum speedup on near-term devices. We employ a self-contained approximate simulator based on NQS methods borrowed from many-body quantum physics, departing from the traditional exact simulations of this class of quantum circuits.

We successfully explore previously unreachable regions in the QAOA parameter space, owing to good performance of our method near optimal QAOA angles. Model limitations are discussed in terms of lower fidelities in quantum state reproduction away from said optimum. Because of such different area of applicability and relative low computational cost, the method is introduced as complementary to established numerical methods of classical simulation of quantum circuits.

Classical variational simulations of quantum algorithms provide a natural way to both benchmark and understand the limitations of near-future quantum hardware. On the algorithmic side, our approach can help answer a fundamentally open question in the field, namely whether QAOA can outperform classical optimization algorithms or quantum-inspired classical algorithms based on artificial neural networks \cite{Gomes2019, Zhao2020, Hibat-Allah2021}.

\section{Methods}
\label{sec:methods}

\subsection{Exact application of one-qubit Pauli gates}
\label{sm:gates}

As mentioned in the main text, some one-qubit gates gates can be applied exactly to the RBM ansatz given in Eq.~\ref{eq:RBM_def}. Here we discuss the specific case of Pauli gates. 
Parameter replacement rules we use to directly apply one-qubit gates can be obtained by solving Eq.~\ref{param_gate_mapping} given in the main text.
Consider for example the Pauli $X_i$ or $\text{NOT} _i$ gate acting on qubit $i$. It can be applied by satisfying the following system of equations:

\begin{equation}
\begin{gathered}
    \ln C + a' _i B_i = (1-B_i) a_i \\
    b' _k + B _i W' _{i k}  =  b_k + (1-B_i) W_{i k} \; .
\end{gathered}
\end{equation}

\noindent for $B_i = 0, 1$. The solution is:
\begin{equation}
\begin{gathered}
    \ln C = a_i \; ; \quad a ' _i = -a_i \; ; \quad \\
    b' _k = b_k + W_{i k} \; ; \quad W' _{i k} = -W_{i k} \; ,
\end{gathered}
\end{equation}

\noindent with all other parameters remaining unchanged.

A similar solution can be found for the Pauli $Y$ gate:
\begin{equation}
\begin{gathered}
    \ln C = a_i + \frac{i \pi}{2} \; ; \quad a ' _i = -a_i + i \pi \; ; \quad \\
    b' _k = b_k + W_{i k} \; ; \quad W' _{i k} = -W_{i k} \; ,
    \end{gathered}
\end{equation}

with all other parameters remaining unchanged as well.

For the Pauli Z gate, as described in the main text, one needs to solve $e^{a' _i B_i} = (-1)^{B_i} e^{a _i B_i}$. The solution is simply
\begin{equation}
    a' _i = a_i + i \pi \; .
\end{equation}

More generally, it is possible to apply exactly an arbitrary $Z$ rotation gate, as given in matrix form as:
\begin{equation}
    RZ(\varphi ) = e^{-i \frac{ \varphi }{2} Z} \propto
    \begin{pmatrix}
        1 & 0\\
        0 & e^{i \varphi}
    \end{pmatrix}
\end{equation}

\noindent where the proportionality is up to a global phase factor. Similar to the Pauli $Z _i$ gate, this gate can be implemented on qubit $i$ by solving $e^{a' _i B_i} = e^{i \varphi B_i} e^{a_i B_i}$. The solution is simply:
\begin{equation}
    a' _i = a_i + i \varphi \; ,
\end{equation}

\noindent with all other parameters besides $a_i$ remaining unchanged. This expression reduces to the Pauli $Z$ gate replacement rules for $\varphi = \pi $ as required.

\subsection{Exact application of two-qubit gates}

We apply two-qubit gates between qubits $k$ and $l$ by adding an additional hidden unit (labeled by $c$) to the RBM before solving Eq.~\ref{param_gate_mapping} from the main text. The extra hidden unit couples only to qubits in question, leaving all previously existing parameters unchanged. In that special case, the equation reduces to

\begin{equation}
\label{eq:two_qubit_eq}
\begin{gathered}
    e^{\Delta a_k B_k + \Delta a_l B_l} \left( 1 + e^{W_{k c} B_k + W_{l c} B_l} \right) \psi_\theta (\mathcal{B}) =\\
    = C \bra{\mathcal{B}} \mathcal{G} \ket{\psi _\theta} \; .
\end{gathered}
\end{equation}

An important two-qubit gate we can apply exactly are ZZ rotations. The gate $RZZ$ is key for being able to implement the first step in the QAOA algorithm. The definition is:
\begin{equation}
    RZZ (\varphi ) = e^{-i \frac{\varphi}{2} Z \otimes Z} \propto
    \begin{pmatrix}
        1 & 0 & 0 & 0 \\
        0 & e^{i \varphi} & 0 & 0 \\
        0 & 0 & e^{i \varphi} & 0 \\
        0 & 0 & 0 & 1
    \end{pmatrix} \; ,
\end{equation}

\noindent where the proportionality factor is again a global phase. The related matrix element for a $RZZ _{k l}$ gate between qubits $k$ and $l$ is $\bra{B' _k B' _l} RZZ _{k l} (\varphi ) \ket{B_k B_l} = e^{i \varphi B_k \oplus B_l}$ where $\oplus$ stands for the classical exclusive or (XOR) operation. Then, one solution to Eq.~\ref{eq:two_qubit_eq} reads:
\begin{equation}
\begin{gathered}
    W_{i c} = -2 \mathcal{A} (\varphi) \; ; \quad
    W_{j c} = 2 \mathcal{A} (\varphi) \\
    a' _i = a_i + \mathcal{A} (\varphi) \; ; \quad
    a' _j = a_j - \mathcal{A} (\varphi) \; ,
\end{gathered}
\end{equation}

\noindent where $\mathcal{A} (\varphi ) = \Arccosh \left( e^{i \varphi} \right)$ and $C=2$.

\subsection{Approximate gate application}
\label{sm:approx_gate}
Here we provide model details and show how to approximately apply quantum gates that cannot be implemented through methods described in Sec.~\ref{sm:gates}.
In this work we use the Stochastic Reconfiguration (SR) \cite{Sorella1998} algorithm to approximately apply quantum gates to the RBM ansatz. To that end, we write the "infidelity" between our RBM ansatz and the target state $\phi$, $\mathcal{D} (\psi _\theta, \phi) = 1 - F(\psi _\theta, \phi)$, as an expectation value of an effective hamiltonian operator $H^\phi _\text{eff}$:
\begin{equation}
\label{eq:fidelity_expectation}
    \mathcal{D}(\psi _\theta, \phi) = \frac{\bra{\psi _\theta} H^\phi _\text{eff} \ket{\psi _\theta}}{\braket{\psi _\theta}{\psi _\theta}} \, \rightarrow \,  H^\phi _\text{eff} = \mathbbm{1} - \frac{ \ketbra{\phi}{\phi}} { \braket{\phi}{\phi} }
\end{equation}

We call the hermitian operator given in Eq.~\ref{eq:fidelity_expectation} a "hamiltonian" only because the target quantum state $\ket{\psi}$ is encoded into it as the eigenstate corresponding to the smallest eigenvalue. Our optimization scheme focuses on finding small parameter updates $\Delta _k$ that locally approximate the action of the imaginary time evolution operator associated with $H ^\phi _\text{eff}$, thus filtering out the target state:
\begin{equation}
\label{eq:imag-time-evol}
    \ket{\psi _{\theta + \Delta} } = C \; e^{-\eta H} \ket{\psi _\theta} \; ,
\end{equation}

\noindent where $C$ is an arbitrary constant included because our variational states (Eq.~\ref{eq:RBM_def}, main text) are not normalized. Choosing both $\eta$ and $\Delta$ to be small, one can expand both sides to linear order in those variables and solve the resulting linear system for all components of $\Delta$, after eliminating $C$ first. After some simplification, one arrives at the following parameter at each loop iteration (indexed with $t$):
\begin{equation}
\label{eq:grad_update}
    \theta ^{(t+1)} _k = \theta ^{(t)} _k - \eta \sum _l S^{-1} _{k l} \; \frac{\partial \mathcal{D}}{\partial \theta ^* _l} \; ,
\end{equation}

\noindent where stochastic estimations of gradients of the cost function $\mathcal{D} (\psi _\theta, \phi)$ can be obtained through samples from $\vert \psi _\theta \vert ^2$ at each loop iteration through:
\begin{equation}
\label{eq:grad_expectation}
    \frac{\partial \mathcal{D}}{\partial \theta _k ^*} = \left\langle \mathcal{O} ^\dagger _k \; H^\phi _\text{eff} \right\rangle _{\psi _\theta} - \left\langle \mathcal{O} ^\dagger _k \right\rangle _{\psi _\theta} \left\langle \vphantom{\mathcal{O} ^\dagger _k } H^\phi _\text{eff} \vphantom{\mathcal{O} ^\dagger} \right\rangle _{\psi _\theta} \; .
\end{equation}

Here, $\mathcal{O} _k$ is defined as a diagonal operator in the computational basis such that $\bra{\mathcal{B} '} \mathcal{O} _k \ket{\mathcal{B}} = \frac{\partial \ln \psi _\theta}{\partial \theta _k} \; \delta _{\mathcal{B} ' \mathcal{B}}$. Averages over $\psi$ are commonly defined as $\langle \cdot \rangle _{\psi} \equiv \nicefrac{\bra{\psi} \cdot \ket{\psi}}{\braket{\psi}}$. Furthermore, the $S$-matrix appearing in Eq.~\ref{eq:grad_update} reads:
\begin{equation}
\label{eq:S_matrix}
     S_{k l} = \left\langle \mathcal{O} ^\dagger _k \mathcal{O} _l \right\rangle _{\psi _\theta} - \left\langle \mathcal{O} ^\dagger _k \right\rangle _{\psi _\theta} \Big\langle \mathcal{O} _l \Big\rangle _{\psi _\theta} \; ,
\end{equation}

\noindent and corresponds to the Quantum Geometric Tensor or Quantum Fisher Information (also see Ref.~\cite{stokes_quantum_2020} for a detailed description and connection with the natural gradient method in classical machine learning \cite{amari_natural_1998}).

Exact computations of averages over $N$ qubit states $\psi _\theta$ and $\phi$ at each optimization step range from impractical to intractable, even for moderate $N$. Therefore, we evaluate those averages by importance-sampling the probability distributions associated with the variational ansatz $\vert \psi _\theta \vert ^2$ and the target state $\vert \phi \vert ^2$ at each optimization step $t$. All of the above expectation values are evaluated using Markov Chain Monte Carlo (MCMC) \cite{Metropolis1953, Hastings1970} sampling with basic single-spin flip local updates. An overview of the sampling method can be found in \cite{Newman1999}. In order to use those techniques, we rewrite Eq.~\ref{eq:grad_expectation} as:
\begin{equation}
\label{eq:param_update}
    \frac{\partial \mathcal{D}}{\partial \theta ^* _k} = \left\langle \frac{\phi}{\psi _\theta} \right\rangle _{\psi _\theta} \left\langle \frac{\psi _\theta}{\phi} \right\rangle _\phi \left[ \left\langle \mathcal{O} ^* _k \right\rangle _{\psi _\theta} - \frac{ \left\langle \frac{\phi}{\psi _\theta} \mathcal{O} ^* _k \right\rangle _{\psi _\theta}}{\left\langle \frac{\phi}{\psi _\theta} \right\rangle _{\psi _\theta}} \right] \; .
\end{equation}

In our experiments with less than 20 qubits, we take 8~000 MCMC samples from 4 independent chains (totaling 32~000 samples) for gradient evaluation. Between each two recorded samples, we take $N$ MCMC steps (for $N$ qubits). For the 54-qubit experiment, we take 2~000 MCMC samples 4 independent chains because of increased computational difficulty of sampling. The entire Eq.~\ref{eq:param_update} is manifestly invariant to rescaling of $\psi _\theta$ and $\phi$, removing the need to ever compute normalization constants. We remark that the prefactor in Eq.~\ref{eq:param_update} is identically equal to the fidelity given in Eq.~\ref{eq:fidelity} in the main text.
\begin{equation}
    F(\psi, \phi) = \frac{\vert \braket{\phi}{\psi} \vert ^2}{\braket{\phi}{\phi} \braket{\psi}{\psi}} = \left\langle \frac{\phi}{\psi} \right\rangle _{\psi} \left\langle \frac{\psi}{\phi} \right\rangle _\phi \; ,
\label{eq:variational_fidelity}
\end{equation}

allowing us to keep track of cost function values during optimization with no additional computational cost.

The second step consists of multiplying the variational derivative with the inverse of the $S$-matrix (Eq.~\ref{eq:S_matrix}) corresponding to a stochastic estimation of a metric tensor on the hermitian parameter manifold. Thereby, the usual gradient is transformed into the natural gradient on that manifold. However, the $S$-matrix is stochastically estimated and it can happen that it is singular. To regularize it, we replace $S$ with $S + \epsilon \mathbbm{1}$, ensuring that the resulting linear system has a unique solution. We choose $\epsilon = 10^{-3}$ throughout. The optimization procedure is summarized in Appendix~\ref{sm:algorithm}.

In order to keep the number of hidden units reasonable, we employ a \textit{compression} step at each QAOA layer (after the first). Immediately after applying the $U_C (\gamma _k)$ gate in layer $k$ to the RBM $\psi _\theta$ (and thereby introducing the unwanted parameters), we go through the following steps:

\begin{enumerate}
    \item Construct a new RBM $\Tilde{\psi} _\theta$.
    \item Initialize $\Tilde{\psi} _\theta$ to exactly represent the state $U_C \left( \frac{1}{k} \sum_{j \leq k} \gamma _j \right) \ket{+}$. Doing this introduces half the number hidden units that are already present in $\psi _\theta$.
    \item Stochastically optimize $\Tilde{\psi} _\theta$ to approximate $\psi _\theta$ (using algorithm in Appendix~\ref{sm:algorithm}) with $\phi \rightarrow \psi _\theta $ and $\psi \rightarrow \Tilde{\psi} _\theta$.
\end{enumerate}

In essence, we use optimization algorithm with the "larger" $\psi _\theta$ as the target state $\phi$. The optimization results in a new RBM state with fewer hidden units that closely approximates the old RBM with fidelity $>0.98$ in all our tests. We then proceed to simulate the rest of the QAOA circuit and apply the same compression procedure again when the number of parameters increases again. The exact schedule of applying this procedure in context of different QAOA layers can be seen on Fig.~\ref{fig:circuit} in the main text.

We choose the initial state for the optimization as an exactly reproducible RBM state that has non-zero overlap with the target (larger) RBM. In principle, any other such state would work, but we heuristically find this one to be a reliable choice across all $p$ values studied. Alternatively, one can just initialize $\Tilde{\psi _\theta}$ to $U_C \left( \gamma ' \right) \ket{+}$ with $\gamma ' = \text{argmax}_\gamma \; F \left(\psi _\theta, U_C (\gamma) \ket{+} \right)$, using an efficient 1D optimizer to solve for $\gamma '$ before starting to optimize the full RBM.

\section*{Data Availability}
The authors declare that the data supporting the findings of this study are available within the paper.

\section*{Code Availability}
Our Python code is available on \href{https://github.com/Matematija/QubitRBM}{GitHub} to reproduce the results presented in this paper through the following URL: \href{https://github.com/Matematija/QubitRBM}{github.com/Matematija/QubitRBM}.

\section*{Acknowledgements}
 The authors thank S.~Bravyi for enlightening discussions and M.~Fishman for insights into MPS simulations. Numerical simulations were performed using NumPy \cite{Harris2020}, SciPy \cite{Virtanen2020}, Google Cirq \cite{Gidney2018} and PastaQ \cite{Torlai2020, Fishman2020} for MPS simulations. Random graph generation was done with NetworkX \cite{Hagberg2008, Steger1999}. Plots were generated using Matplotlib \cite{Hunter2007}. MM acknowledges support from the CCQ graduate fellowship in computational quantum physics. The Flatiron Institute is a division of the Simons Foundation.

\section*{Author contribution statement}
G.C. conceived the main idea and co-wrote the manuscript. M.M. developed the idea further, wrote the computer code, executed the numerical simulations and co-wrote the manuscript.

\section*{Competing Interest Statement}
The authors declare no Competing Financial or Non-Financial Interests.

\bibliographystyle{naturemag}
\bibliography{references}

\onecolumngrid
\appendix

\section{Exact p=1 formula}
\label{sm:formula}

In this section we present the exact MaxCut QAOA cost formula used for quantifying the accuracy of our model result.

\begin{theorem}
\label{thm:formula}
For an arbitrary graph $G$, the QAOA cost function for the MaxCut problem given in Eq.~\ref{eq:qaoa_cost_general} (main text) takes the form

\begin{equation}
\label{eq:theorem_formula}
    C(\gamma, \beta) = \frac{1}{2} \sum _{\langle k ,l \rangle} \left[
     \vphantom{\frac{1}{2}} \sin (4 \beta) \sin (2 \gamma) \left( \cos ^{q_k} (2 \gamma ) + \cos ^{q_l} (2 \gamma ) \right) + \sin ^2 (2 \beta ) \cos ^{q_k + q_l - 2 \Delta _{k l}} (2 \gamma) (1- \cos ^{\Delta_{k l}} (4 \gamma)) \right]
\end{equation}

\noindent at $p=1$. Here, $q_k + 1$ and $q_l +1$ are degrees of vertices $k$ and $l$ and $\Delta _{k l}$ is the number of common neighbors between those vertices.
\end{theorem}

\textbf{Proof:} The proof repeats much of what has already been done in \cite{Wang2018}. We begin by expressing the density operator associated with the $\ket{+}$ state as

\begin{equation}
    \rho _0 = \ketbra{+}{+} = \prod _i \frac{\mathbbm{1}_i + X_i}{2} \; .
\end{equation}

Then, we can express the MaxCut QAOA cost function at $p=1$ as

\begin{equation}
    \bra{\gamma, \beta} \mathcal{C} \ket{\gamma, \beta} = \sum _{\langle k, l \rangle } \Tr \left[ \rho _0 U_C ^\dagger (\gamma ) U_B ^\dagger (\beta ) Z_k Z_l U_B (\beta ) U_C (\gamma )  \right]
\end{equation}

In what follows, we will make repeated use of the following identities:
\begin{align}
    e^{i \beta X} Z e^{-i \beta X} &= \cos (2 \beta ) \; Z + \sin (2 \beta ) \; Y \label{eq:id1} \\
    e^{-i \gamma Z \otimes Z} &= \cos \gamma - i \sin \gamma \; Z \otimes Z \label{eq:id2} \\
    e^{-i \gamma Z} \; Y &= Y \; e^{i \gamma Z} \label{eq:id3} \; .
\end{align}

The innermost product can easily be expanded using Eq.~\ref{eq:id1} twice:

\begin{equation}
\label{eq:proof_3_terms}
\begin{gathered}
    U_B ^\dagger (\beta ) Z_k Z_l U_B (\beta ) = U_B ^{\dagger} (\beta ) Z_k U_B (\beta ) U_B ^{\dagger} (\beta ) Z_l U_B (\beta ) = \\
    = \cos ^2 (2 \beta ) \; Z_k Z_l + \frac{1}{2} \sin (4 \beta ) \left( Y_k Z_l + Y_l Z_k \right) + \cos ^2 (2 \beta ) \; Y_k Y_l \; .
\end{gathered}
\end{equation}

The first term vanishes when averaged against $\rho _0$. The second and third need to be treated separately. First we look at expressions of the form:

\begin{equation}
\label{eq:proof_term_1}
    \Tr \left[ \rho _0 \; e^{i\gamma \mathcal{C}} \; Y_k Z_l \;  e^{-i\gamma \mathcal{C}} \right] = \Tr \left[ \rho _0 \; Y_k Z_l \prod _{j \in N(k)} e^ {-2 i \gamma Z_k Z_j} \right] = \Tr \left[ \rho _0 \; Y_k Z_l \prod _{j \in N(k)} \left( \cos 2 \gamma - i \sin 2 \gamma \; Z_k  Z _j \right) \right] \, ,
\end{equation}

\noindent where we denoted the set of all neighbors of node $k$ by $N(k)$, used Eq.~\ref{eq:id3} and Eq.~\ref{eq:id2}. In Eq.~\ref{eq:proof_term_1}, tracing out Paulis not associated to either node $k$ or its neighbors is trivial and produces a factor of unity. Furthermore, tracing out immediate neighbors of $k$ other than $l$ produces a factor of $\cos ^{q_k} (2 \gamma )$ where $q_k + 1$ is the degree of node $k$. Keeping those factors aside, we are left with:

\begin{equation}
    \Tr _{k l} \left[ \frac{\mathbbm{1}_k + X_k}{2} \frac{\mathbbm{1}_l + X_l}{2} \; Y_k Z_l \left( \cos 2\gamma - i \sin 2 \gamma \; Z_k  Z _l \right) \right] = \sin 2\gamma \; .
\end{equation}

Therefore, the final contribution from the second term in Eq.~\ref{eq:proof_3_terms} is

\begin{equation}
    \frac{1}{2} \sin (4 \beta)  \Tr \left[ \rho _0 \; e^{i\gamma \mathcal{C}} \left( Y_k Z_l + Y_l Z_k \right) \; e^{-i\gamma \mathcal{C}} \right] = \frac{1}{2}  \sin (2\gamma ) \sin (4 \beta ) \left[ \cos ^{q_k} (2 \gamma ) + \cos ^{q_l} (2 \gamma ) \right]
\end{equation}

Looking at the last factor in Eq.~\ref{eq:proof_3_terms}, we compute:

\begin{equation}
\label{eq:proof_term_2}
\begin{gathered}
    \Tr \left[ \rho _0 \; e^{i\gamma \mathcal{C}} Y_k Y_l e^{-i\gamma \mathcal{C}} \right] = \Tr \left[ \rho _0 \; Y_k Y_l \; \left( \prod _{\substack{j \in N(k) \\ j \neq l}} e^{-i \gamma Z_k Z_j} \right) \left( \prod _{\substack{i \in N(l) \\ i \neq k}} e^{-i \gamma Z_k Z_j} \right) e^{i \gamma Z_k Z_l} e^{-i\gamma \mathcal{C}} \right] = \\
    = \Tr \left[ \rho _0 \; Y_k Y_l \; \left( \prod _{\substack{j \in N(k) \\ j \neq l}} e^{-2 i \gamma Z_k Z_j} \right) \left( \prod _{\substack{i \in N(l) \\ i \neq k}} e^{-2 i \gamma Z_k Z_j} \right) \right] \; ,
\end{gathered}
\end{equation}

\noindent where we separate the factor in $\mathcal{C}$ corresponding to the edge $(k,l)$ out before using Eq.~\ref{eq:id3}. In the second equality, the factor corresponding to the edge $(k, l)$ cancels and others are absorbed into product expressions.

Like before, operators corresponding to nodes other than $k$ or $l$ or their neighbors can be traced out trivially. The next step is to trace out Paulis that are neighbors of $k$ but not $l$. Each of those contributes with a factor of $\cos (2 \gamma)$ so we get a factor of $\cos ^{q_k - \Delta _{k l}}  (2 \gamma)$ for node $k$ and a corresponding factor for node $l$ resulting in $\cos ^{q_k + q_l - 2 \Delta _{k l}}  (2 \gamma)$. Here, we denoted the number of common neighbors of $k$ and $l$ by $\Delta _{k l}$ and label their set by $CN(k,l)$ in what remains of Eq.~\ref{eq:proof_term_2}:

\begin{equation}
\label{eq:proof_term_2_2}
    \Tr \left[ \rho _0 \; e^{i\gamma \mathcal{C}} Y_k Y_l e^{-i\gamma \mathcal{C}} \right] = \cos ^{q_k + q_l - 2 \Delta _{k l}}  (2 \gamma) \; \Tr \left[ \rho ' _0 \;  Y_k Y_l \prod _{\substack{ i \in CN(k, l)}} \left( e^{-2 i \gamma Z_k Z_i} e^{-2 i \gamma Z_l Z_i} \right) \right] \; ,
\end{equation}

\noindent where $\rho ' _0$ is the reduced density operator. Using Eq.~\ref{eq:id2}, one can trace out nodes in $CN(k,l)$ to obtain a factor of $(\cos ^2 2\gamma + \sin ^2 2\gamma \; Z_k Z_l)^{\Delta _{k l}}$. Finally, we expand that expression using the binomial theorem and interchange the trace operation with the sum. The trace in Eq.~\ref{eq:proof_term_2_2} becomes

\begin{equation}
    \cos ^{2 \Delta _{k l} } (2 \gamma) \; \sum _{\substack{n=0 \\ \text{odd } n}} ^{\Delta _{k l}} \binom{\Delta _{k l} }{n} \tan ^{2n} (2 \gamma ) = \frac{1}{2} \left( 1 - \cos ^{\Delta _{k l}} (4 \gamma ) \right)
\end{equation}

Putting all of the ingredients together, we have:

\begin{equation}
    C(\gamma, \beta) = \frac{1}{2} \sum _{\langle k ,l \rangle} \left[ \vphantom{\frac{1}{2}} \sin (4 \beta) \sin (2 \gamma) \left( \cos ^{q_k} (2 \gamma ) + \cos ^{q_l} (2 \gamma ) \right) + \sin ^2 (2 \beta ) \cos ^{q_k + q_l - 2 \Delta _{k l}} (2 \gamma) (1- \cos ^{\Delta_{k l}} (4 \gamma)) \right]
\end{equation}

\section{Algorithm for stochastic gate application}
\label{sm:algorithm}

\begin{algorithm*}
    \SetAlgoLined
    \KwIn{Initial parameters $\theta $, target state $\phi$, target fidelity tolerance \texttt{tol}, learning rate $\eta$}
    \KwResult{Parameters $\theta '= \{ \vb{a}', \vb{b}', W' \}$ such that $\ket{\psi _{\theta '}} \approx \ket{\phi}$}
    $\partial _k \ln F \gets 0 \; ; S_{k l} \gets 0 \; ; \theta ' \gets \theta $ \;
    \While{$F < 1 - \texttt{tol}$}{
    Generate samples $\mathcal{B} _\psi \sim \vert \psi _{\theta '} \vert ^2$ and $\mathcal{B} _\phi \sim \vert \phi \vert ^2$  using MCMC\;
    \lForEach{$\mathcal{B}$ in $\mathcal{B} _\psi$}{
    evaluate and store $\psi _{\theta '} (\mathcal{B} )$, $\phi (\mathcal{B})$ and $\mathcal{O}_k (\mathcal{B})$ for all parameters
    }
    \lForEach{$\mathcal{B}$ in $\mathcal{B} _\phi$}{
    evaluate and store $\psi _{\theta '} (\mathcal{B} )$ and $\phi (\mathcal{B})$
    }
    evaluate and set $F \gets \Re \left\{
        \left\langle \frac{\phi (\mathcal{B})}{\psi (\mathcal{B})} \right\rangle _{\mathcal{B} \sim \mathcal{B} _\psi} \left\langle \frac{\psi (\mathcal{B})}{\phi (\mathcal{B})} \right\rangle _{\mathcal{B} \sim \mathcal{B} _\phi}
         \right\} $ \;
    \lForEach{$k$}{evaluate and set $\partial _k \ln F \gets
        \left\langle \mathcal{O} ^* _k \right\rangle _{\mathcal{B} \sim \mathcal{B} _\psi} - \left\langle \frac{\phi}{\psi _\theta} \mathcal{O} ^* _k \right\rangle _{\mathcal{B} \sim \mathcal{B} _\psi} \left/ \left\langle \frac{\phi}{\psi _\theta} \right\rangle _{\mathcal{B} \sim \mathcal{B} _\psi} \right. $}
    \lForEach{$k, l$}{evaluate and set $S_{k l} \gets \left\langle \mathcal{O} ^\dagger _k \mathcal{O} _l \right\rangle _{\psi _\theta} - \left\langle \mathcal{O} ^\dagger _k \right\rangle _{\psi _\theta} \left\langle \vphantom{\mathcal{O} ^\dagger} \mathcal{O} _l \right\rangle _{\psi _\theta}$}
    solve linear system $ \sum _l S_{k l} \Delta _l = F \times \partial _k \ln F $ \;
    \lForEach{$k$}{$\theta ' _k \gets \theta ' _k - \eta \; \Delta _k$}
    }
    \Return $\theta '$
    \caption{Target state approximation with an RBM.}
    \label{alg:SR}
\end{algorithm*}

\section{Relation to t-VMC}
\label{sm:tvmc}

Time-dependent Variational Monte Carlo (t-VMC) \cite{carleo_localization_2012} is a numerical technique used in many-body quantum physics to approximately capture time evolution of an arbitrary state often captured by the unitary operator $e^{-i H t}$ associated with the system hamiltonian $H$. The starting point of such calculations is often almost identical to Eq.~\ref{eq:imag-time-evol}:

\begin{equation}
    \ket{\psi _{\theta (t + \Delta t)}} = C \; e^{-i H \Delta t} \ket{\psi _{\theta (t)}} \; ,
\end{equation}

\noindent after defining an appropriate variational ansatz $\psi _\theta$, usually by defining log-derivative operators $\mathcal{O} _k$, like in the main text of this work. By repeating the calculation outlined in the previous subsection (which in this case is equivalent to plugging into the time-dependent Schrödinger equation), one obtains the so-called optimal equations of motion:

\begin{equation}
    \sum _l \langle \mathcal{O} ^* _k \mathcal{O} _l \rangle ^c _t \; \Dot{\theta } _k (t) = -i \langle \mathcal{O} ^* _k H \rangle ^c _t \; ,
\end{equation}

\noindent where $\langle A B \rangle ^c _t \equiv \langle A B \rangle _t - \langle A \rangle _t \langle B \rangle _t$ and $\langle \cdots \rangle _t \equiv \bra{\psi _\theta (t)} \cdots \ket{\psi _\theta (t)}/ \braket{\psi _\theta (t)}{\psi _\theta (t)} $. At this point, any standard ODE solver can be employed to propagate parameters $\theta _k$ away from the initial condition after using MCMC to stochastically estimate averages $\langle \cdots \rangle ^c _t$ at each step. We note that t-VMC requires inverting the matrix $\langle \mathcal{O} ^* _k \mathcal{O} _l \rangle ^c _t$ which is analogous to the $S$ matrix given in Eq.~\ref{eq:S_matrix} in the main text.

In the case of QAOA, one might wish to approximate the action of $U_B (\beta ) = \exp \left( -i \beta \sum _j X_j \right)$ by employing t-VMC to "propagate" the relevant parameters in $\beta$ instead of physical time. That approach would have the benefit of being able to apply the entire $U_B$ gate in small $\Delta \beta$ increments instead of doing it qubit-by-qubit. Indeed, if we take Eq.~\ref{eq:param_update} from the main text and set $\ket{\phi} = e^{-i \Delta \beta X_j} \ket{\psi _\theta}$, we obtain:

\begin{equation}
    \frac{\partial \mathcal{D}}{\partial \theta ^* _k} = \left( \frac{i}{2} \sin (2 \Delta \beta) - \sin ^2 (\Delta \beta ) \left\langle \frac{ \psi ^{X_j} _\theta}{\psi _\theta} \right\rangle ^* _{\psi _\theta} \right) \left[ \left\langle \mathcal{O} ^* _k \frac{ \psi ^{X_j} _\theta}{\psi _\theta} \right\rangle _{\psi _\theta} - \left\langle \mathcal{O} ^* _k \right\rangle _{\psi _\theta} \left\langle \frac{ \psi ^{X_j} _\theta}{\psi _\theta} \right\rangle _{\psi _\theta} \right] \; ,
\end{equation}

\noindent where $\psi ^{X_j} _\theta (\mathcal{B}) \equiv \bra{\mathcal{B}} X_j \ket{\psi _\theta}$. If one expands the first factor to first order in $\Delta \beta$, what is left is exactly the t-VMC parameter update one would obtain by following the t-VMC derivation from the beginning. Therefore, our method is more general than t-VMC for any individual qubit. In addition to generality, it has two other key features:

\begin{enumerate}
    \item Fidelity $F(\phi, \psi)$ is directly used as a cost function and it is recorded at each optimization step. This provides for a more controlled environment where fidelity is explicitly optimized over rather than implicitly. We can apply as many gradient updates as needed to reach the target fidelity.
    \item The computational cost of complete (approximate) application of $U_B$ is similar between the two methods. We report that the explicit fidelity method we used in this work required approximately 30 updates per qubit while t-VMC needed 1000-2000 intermediate $\beta$ points to reach similar final fidelities on 20-qubit test systems.
\end{enumerate}

\end{document}